\documentclass[epj]{webofc}
\usepackage[varg]{txfonts}   
\woctitle{MESON2016 - the 14$^\textrm{th}$ International Workshop on Meson Production, Properties and Interaction}
\begin{document}
\selectlanguage{english}
\title{Study of the lowest tensor and scalar resonances
\\  in the $\tau \to \pi\pi\pi \nu_\tau$ decay}

\author{Olga~Shekhovtsova\inst{1,2}\fnsep\thanks{Speaker, \email{oshekhov@ifj.edu.pl}}\fnsep\thanks{IFJPAN-IV-2016-22 } \and
        Juan~Jos\'e~Sanz-Cillero \inst{3} \and
        T.~Przedzi\'nski\inst{4} 
}

\institute{Institute of Nuclear Physics PAN,
 Cracow, Poland
\and
           NSC KIPT Institute for Theoretical Physics,  Kharkov, Ukraine 
\and
          Departamento de F\'isica Te\'orica and Instituto de F\'isica Te\'orica,
IFT-UAM/CSIC, Universidad Aut\'onoma de Madrid, Cantoblanco, 28049 Madrid, Spain          
\and
The Faculty of Physics, Astronomy and Applied Computer Science, Jagellonian   University,  Cracow, Poland}

\abstract{
  In this note we present a new parametrization of the hadronic current for the decay $\tau \to \pi\pi\pi \nu_\tau$  derived from the chiral lagrangian with explicit inclusion of resonances. We have included both scalar, vector and axial-vector resonances. For the first time, the lowest tensor resonance ($f_2(1270)$) is included as well.  Both single and double-resonance contributions to the hadronic form factors are taken into account.  To satisfy the correct high energy behaviour of the hadronic form factors, constraints on numerical values of the vertex constants are obtained.
}
\maketitle
\section{Introduction}
\label{intro}
Hadronic decay modes of $\tau$-lepton gives information about the hadronization mechanism and resonance dynamics in the energy region, where the pQCD methods are not applicable. 
In the last years substantial progress for the simulation of the process $\tau \to 3\pi \nu_\tau$ was achieved.
 The progress~\cite{Nugent:2013hxa} was related to a new parametrization of the hadronic current based on the Resonance Chiral Lagrangian (RChL)  and to the recent availability
of the unfolded distributions from preliminary BaBar analysis~\cite{Nugent:2013ij} for all invariant
hadronic masses for the three-prong mode. The lowest-energy scalar resonance was added phenomenologically and, as a result, the corresponding hadronic current does not reproduce the correct chiral low-energy behaviour and the $\pi^0\pi^0\pi^-$ and $\pi^-\pi^-\pi^+$ amplitudes do not reproduce the isospin relation~\cite{Girlanda:1999fu}. Comparison with the data has demonstrated also a hint on the missing tensor resonance ($f_2(1270)$).

The goal of this note is to outline a consistent model to describe the tau-lepton decys into three pions based on RChL with scalar  ($J^{PC}=0^{++}$) and tensor resonances  ($J^{PC}=2^{++}$) and that fulfill the high-energy QCD and low-energy chiral limits for the hadronic form-factors. The detail description of the model and calculation of the hadronic form-factor will be presented~\cite{jj_and_olga}.

\section{Three pion hadronic current. Axial-vector form-factors related with scalar and tensor resonances.}
The most general Lorenz invariant current for 
$\tau^-\to \pi^{-(0)}(p_1)\pi^{-(0)}(p_2)\pi^{-(+)}(p_3) \nu_\tau$
\begin{equation}
\!\!H_\mu^{3\pi}(p_1,p_2,p_3)= 
iP_T(q)^{\alpha}_{\,\mu} \left(
(p_1^\mu -p_3)^\mu \mathcal{F}_2(s_1,s_2,q^2)
+
(p_2^\mu -p_3)^\mu \mathcal{F}_1(s_1,s_2,q^2)
\right)+ iq^\alpha\,\mathcal{F}_P(s_1,s_2,q^2) \nonumber
\end{equation}
with $s_i= (p_j-p_k)^2  , \, \, \, q^2 = (p_1+p_2+p_3)^2$ and, due to the Boson symmetry, the hadronic form-factors are related:
$\mathcal{F}_2(s_1,s_2,q^2) = \mathcal{F}_1(s_2,s_1,q^2)$.
The longitudinal form-factor $\mathcal{F}_P$ is suppressed by $m_\pi^2/q^2$ compared to $\mathcal{F}_{1,2}$ and in this note we will neglect it.
\begin{figure}[ht]
\centering
\vspace{-0.2cm}
\includegraphics[width = 0.75\textwidth,height = 0.12\textwidth]{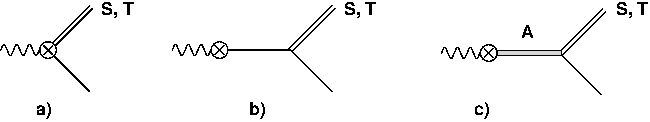}
\vspace{-0.4cm}
\caption{Relevant diagrams for the $\tau$-lepton decay into $S \pi$ and $T \pi$. Single straight lines are for pions, the wave line is for an incoming $W^-$.}\label{fig:diagr}
\vspace{-0.5cm}
\end{figure}
In Fig.~\ref{fig:diagr}, we show the three relevant diagrams that must be taken into account: a) the direct production; b) the intermediate $\pi^-$ production and c) the double resonance production through the intermediate $a_1$ axial-vector resonance. To calculate the corresponding diagrams we use the RChL approach~\cite{Ecker:1988te} for the vector and axial-vector ($A$) resonances combined wothw the Lagrangian including interaction of a tensor ($T$) multiplet and pions~\cite{Zauner:2007}. Moreover we add the operators with two resonances:
\begin{itemize}
\item $A S\pi$ interaction
$\Delta\mathcal{L}_{AS\pi} = \lambda_1^{AS} \langle \{\nabla_\mu S , A^{\mu\nu}\} u_\nu\rangle $
\item $A T\pi$ interaction
$\Delta \mathcal{L}_{AT\pi} = \lambda_1^{AT\pi} \langle \{ A_{\alpha\beta} , \nabla^\alpha T_{\, \mu}^\mu \}  u^\beta \rangle
+ \lambda_2^{AT\pi} \langle \{ A_{\alpha\beta} , \nabla_\mu T^{\mu\alpha}\} u^\beta \rangle \,  $,
\end{itemize}
where for the axial-vector field $A_{\alpha\beta}$ we apply the antisymmetric tensor representation~\cite{Ecker:1988te}, $S$ is the scalar field, a tensor multiplet is $T^{\mu\nu}=f_2^{\mu\nu}/\sqrt{2}*diag(1,1,0)$  (we will assume the ideal mixing in the tensor nonet and that the $f_2(1270)$ resonance is pure $u\bar{u}+d\bar{d}$). 

The $\pi^0\pi^0\pi^-$ and $\pi^-\pi^-\pi^+$ amplitudes obey the isospin  relation~\cite{Girlanda:1999fu} 
that leads to
\begin{eqnarray}\label{eq.ff-rel}
\mathcal{F}_1^{--+}(s_1,s_2,q^2) =  \mathcal{F}_1^{00-}(s_3,s_2,q^2)-\mathcal{F}_1^{00-}(s_3,s_1,q^2)-\mathcal{F}_1^{00-}(s_1,s_3,q^2) \, . 
\end{eqnarray}

For the three-pion form-factor caused by the intermediate $\sigma$-resonance we have:
\begin{eqnarray}\label{eq.ff-scal}
\mathcal{F}_1(s_1,s_2,q^2)^{00-} &=&  \frac{2}{3} \mathcal{F}_{S\pi}^a(q^2 ;s_3  )\, \mathcal{G}_{S \pi\pi}(s_3)\, ,
\end{eqnarray}
where  the $A S\pi$ form-factor and propagation of the $\sigma$-resonance and its decay into $\pi\pi$ are
\begin{eqnarray}
\mathcal{F}^a_{S\pi}(q^2  ;k^2   )&=& \frac{2c_d}{ F_\pi  }
\quad + \quad
\frac{\sqrt{2} F_A \lambda_1^{AS}}{  F_\pi  }\frac{q^2}{M_A^2-q^2} \, , \,\, 
\mathcal{G}_{S\pi\pi}(s_3)= \frac{\sqrt{2}c_d}{ F_\pi^2 } \,\frac{(s_3-2m_\pi^2)}{M_S^2 -     s_3     } \nonumber
\end{eqnarray}
and  $qp_j = (m_\pi^2 +q^2 -s_j)/2$.  Requiring  $\mathcal{F}^a_{S\pi}(q^2;k^2) \rightarrow 0$ for $q^2 \rightarrow \infty$ we got $F_A\lambda_1^{AS} = \sqrt{2}c_d$. 

To include 
a $\sigma$--$f_0(980)$ splitting and non-zero width of the resonances
we follow~\cite{Escribano:2010wt}
\begin{equation}
\frac{1}{M_S^2\,-\,s}  \qquad  \longrightarrow \qquad
\frac{\cos^2\phi_S}{
M_\sigma^2\,-\,s  -\, f_{\sigma}(s)\, -\, i M_\sigma \Gamma_\sigma(s) 
 } \, \,+\,\, \frac{\sin^2\phi_S}{M_{f_0}^2\,-\,s - \, i M_{f_0} \Gamma_{f_0}} \, ,
 \label{eq.MS-splitting}
\end{equation}
where $\phi_S$ is the scalar mixing angle. For the $f_0$ parameters we will use the numerical values $M_{f_0}=980$~MeV,
$\phi_S=-8^\circ$~\cite{Escribano:2010wt}. As a first approach we also consider  the Breit-Wigner function for the $\sigma$-propagator in our numerical study. 

Schematically the form-factor related with the intermediate tensor resonance state is written as
\begin{eqnarray}\label{eq.ff-tens}
\! \! \!\mathcal{F}_1(s_1,s_2,q^2)^{00-} &=& \frac{H_1(q^2,s_1,s_2)}{(M_A^2-q^2)(M_{f_2}^2-s_3)} + \frac{H_2(q^2,s_1,s_2)}{(M_A^2-q^2)} + \frac{H_3(q^2,s_1,s_2)}{(M_{f_2}^2-s_3)} + H_4(q^2,s_1,s_2) \, ,
\end{eqnarray}
where $H_i(q^2,s_1,s_2)$ are non-singular functions. 
We would like to stress that for $q^2 = M_A^2$ and $s_3 = M_{f_2}^2$ our expression (\ref{eq.ff-tens}) reproduces the corresponding contribution of Eq.~(A.3) of~\cite{Asner:1999kj} and that in~\cite{Castro:2011zd}. However, for an arbitrary off-shell momentum of the intermediate tensor resonance we have
a more general momentum structure of the hadronic current, which also ensures
the right low energy behaviour and the transversality of the matrix element in the chiral limit. As a result it brings three additional functions $H_{2,3,4}(q^2,s_1,s_4)$ in (\ref{eq.ff-tens}) (see for discussion~\cite{jj_and_olga}).

To obtain the $\pi^-\pi^-\pi^+$ form-factors we apply the relation~(\ref{eq.ff-rel})  for~(\ref{eq.ff-scal}) and (\ref{eq.ff-tens}). Exact formulae are presented in~\cite{jj_and_olga}.

\begin{figure}[ht]
\vspace{-0.2cm}
\centering
\includegraphics[width = 0.85\textwidth,height = 0.23\textwidth]{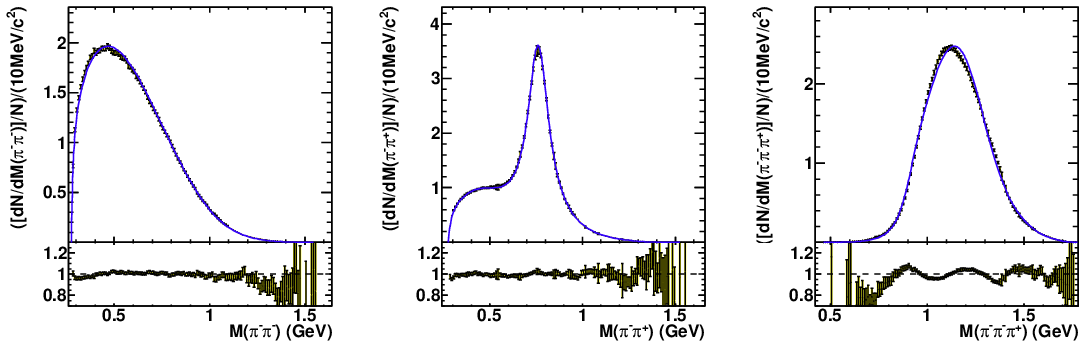}
\vspace{-0.4cm}
\caption{The $\tau^-\to \pi^-\pi^-\pi^+\nu_\tau$ decay invariant mass distribution. The preliminary BaBar data~\cite{Nugent:2013ij} are presented by points and the line corresponds to the model.}\label{fig:fit}
\vspace{-0.6cm}
\end{figure}

The hadronic form-factors~(\ref{eq.ff-scal}) and~(\ref{eq.ff-tens}) have been implemented in the Monte Carlo Tauola~\cite{Nugent:2013hxa}. To get the model parameters the one-dimentional spectra $d\Gamma/ds_1$, $d\Gamma/ds_3$ and $d\Gamma/dq^2$ with the hadronic form-factors~(\ref{eq.ff-scal}) and~(\ref{eq.ff-tens})  in addition to~\cite{Shekhovtsova:2012ra} have been fitted to the preliminary $\pi^-\pi^-\pi^+$ BaBar data~\cite{Nugent:2013ij}. The results are presented in Fig.~\ref{fig:fit} (as an example we present the result for the Breit-Wigner $\sigma$-meson propagator). For the first approach we have fixed the tensor resonance parameters to their PDG values. The difference between the data and the theoretical distributions is less than $5-7\%$,  except for the low- and high-energy tails, where the statistics is low.  
The inclusion of the tensor resonance contribution in the fit and the study of the fit stability and systematic uncertainities are in progress.

\begin{acknowledgement}
The work of J.J.S.C is partially supported by grant FPA2013-44773-P
and the Centro de Excelencia Severo Ochoa Programme  (Spanish Ministry
MINECO) SEV-2012-0249, the research of O.Sh. was supported in part by funds of the Foundation of Polish Science grant POMOST/2013-7/12.
\end{acknowledgement}
%
%
%

\end{document}